\documentclass[]{emulateapj}
\usepackage[breaklinks,colorlinks,citecolor=blue,linkcolor=magenta]{hyperref} 
\usepackage{amsgen,amsmath,amstext,amsbsy,amsopn,amssymb}
\usepackage{epstopdf}
\usepackage{graphics}
\usepackage[usenames]{color}
\usepackage{epsfig}
\usepackage{longtable}
\usepackage{bm} 
\usepackage{ulem}
\usepackage{url}
\usepackage{fontenc}
\usepackage{inputenc}
\usepackage{xcolor}
\usepackage{varwidth}
\usepackage{braket}

\pdfoutput = 1

\begin{document}
\raggedbottom

\preprint{}

\title{Probing Cosmic Origins with CO and [CII] Emission Lines}

\author{Azadeh Moradinezhad Dizgah$^{1}$$^{\ast}$, Garrett K. Keating$^{2}$, and Anastasia Fialkov$^{2,3,4,5}$}
\affil{\small $^{1}$Department of Physics, Harvard University, 17 Oxford St., Cambridge, MA 02138, USA}
\affil{\small $^{2}$Harvard-Smithsonian Center for Astrophysics, 60 Garden Street, Cambridge, MA 02138, USA}
\affil{\small $^{3}$Institute of Astronomy, University of Cambridge, Madingley Road, Cambridge CB3 0HA, UK}
\affil{\small $^{4}$Kavli Institute for Cosmology, University of Cambridge, Madingley Road, Cambridge CB3 0HA, UK}
\affil{\small $^{5}$Department of Physics, The University of Tokyo, 7-3-1 Hongo, Bunkyo, Tokyo 113-0033, Japan}
\email{$^{\ast}$amoradinejad@physics.harvard.edu}

\begin{abstract}
Primordial non-Gaussianity is an invaluable window into the physical processes that gave rise to cosmological structure.  The presence of local shape PNG imprints a distinct scale-dependent correction to the bias of dark matter tracers on large scales, which can be effectively probed via the technique of intensity mapping. Considering an upcoming generation of experiments, we demonstrate that intensity mapping of CO and [CII] emission can improve upon the current best constraints from the {\it Planck} satellite. We show that measurement of the CO intensity power spectrum by a hypothetical next stage of the ground-based COMAP experiment can achieve  $\sigma(f_{\rm NL}^{\rm loc}) = 3.4$, and that the proposed CMB satellite mission PIXIE can achieve $\sigma(f_{\rm NL}^{\rm loc}) = 3.9$ via measurement of [CII] intensity power spectrum.
\end{abstract}


\maketitle
\section{Introduction}
Understanding the origin of structure in the universe is a key open question in cosmology. Inflation is the leading paradigm of the early universe, in which quantum fluctuations of a scalar field, i.e. inflaton, planted the seeds for the formation of the structure. The simplest models of inflation, characterized by a canonical single scalar field, originating from the Bunch-Davies vacuum and slowly rolling down its potential, predict a nearly Gaussian distribution of primordial perturbations. Deviations from these simple models can produce a distinctive, non-Gaussian signature which, to leading order, results in non-zero bispectrum of primordial curvature perturbations $\zeta$ (see \cite{Bartolo:2004if, Chen:2010xka} for reviews). It is common to write the bispectrum in terms of a scale-independent amplitude $f_{\rm NL}$ and a shape function; constraint of this amplitude (for a given shape) is a unique probe to discriminate between models of inflation.

Primordial non-Gaussianity (PNG) of the local shape \citep{Gangui:1993tt,Wang:1999vf, Verde:1999ij, Komatsu:2001rj}, which is produced by super-horizon, non-linear evolution of $\zeta$, can be parametrized by a nonlinear correction to the Gaussian perturbations $\zeta_G$, as $\zeta = \zeta_G + 3/5 f_{\rm NL}^{\rm loc} (\zeta_G^2 - \langle \zeta_G ^2\rangle)$. The local shape PNG is a sensitive probe of multi-field inflation, as single-field scenarios, in the attractor regime, are expected to produce $f_{\rm NL}^{\rm loc} \ll 1$ \citep{Maldacena:2002vr, Creminelli:2004yq}.  Current best constraints on $f_{\rm NL}^{\rm loc}$ are set by measurements of the cosmic microwave background (CMB) from the {\it Planck} satellite \citep{2016A&A...594A..13P}. These results are consistent with Gaussian primordial fluctuations ($f_{\rm NL}^{\rm loc}{=}0$), with a 1$-\sigma$ uncertainty -- when translated to large-scale structure (LSS) conventions -- of $\sigma(f_{\rm NL}^{\rm loc}) \simeq 6.5$ \citep{Camera:2014bwa}. These constraints can be significantly improved by measurements of the statistical properties of LSS (see \citep{2014arXiv1412.4671A} for an overview). Among other effects, PNG leaves an imprint on the power spectrum of biased tracers of dark matter by inducing a scale-dependent correction to their bias, which is significant on large scales \citep{Dalal:2007cu,Matarrese:2008nc,Afshordi:2008ru}. In this Letter, we show that with this signature, power spectrum measurements from intensity mapping of CO and [CII] line emission have the potential to improve the constraints on PNG beyond the current best limits.

In contrast to galaxy surveys, which aim to detect groups of individual sources to some threshold significance and completeness, line intensity mapping probes the large-scale matter distribution by measuring the cumulative light from an ensemble of sources, including faint, unresolved galaxies, while preserving accurate redshift information. Previous studies have shown that intensity mapping of the 21-cm line of neutral hydrogen (HI) at redshifts  $z=1-5$,  with purpose-designed surveys, can produce constraints of order  $\sigma(f_{\rm NL}^{\rm loc})\sim 1$  \citep{Camera:2013kpa}. Interest in other emission lines as candidates for intensity mapping has been bolstered by the tentative power spectrum detections of CO and [CII] \citep{2016ApJ...830...34K,2018MNRAS.478.1911P}, and the multitude of upcoming intensity mapping surveys (see \citep{2017arXiv170909066K} for a recent summary). Here, we provide the first forecast for the potential of such surveys in constraining PNG, considering experimental setups targeting CO and [CII] emission from as far back in time as the Epoch of Reionization (EoR; $z\sim6{-}10$), mapping the cosmic web at redshifts and scales that are inaccessible to upcoming spectroscopic/photometric galaxy surveys. 

\section{The line intensity power spectrum} 
CO is predominantly found in the dense clouds of molecular gas (${\rm of \ density} \ n \sim10^{3}$ cm$^{-3}$), while [CII] is found in the neutral media of galaxies ($n \sim1$ cm$^{-3}$) \citep{Carilli:2013qm}. Both are typically tracers of the cold gas within galaxies that provides the fuel for star formation, and the strength of their emission is observed to be correlated with the star formation rates of galaxies \citep{2013ApJ...768...74T, Herrera-Camus:2015}. Under the assumption that line emission for both CO and [CII] arise primarily from within galaxy host halos, and that the luminosities of these lines can be expressed as a function of halo mass, the mean brightness temperature (typically in units of $\mu K$) can be written as 
\begin{equation}
\langle T_{\rm line}\rangle (z)  = \frac{c^2 }{2k_B \nu_{\rm obs}^2} {\int} dM \frac{dn}{dM} \frac{L(M,z)}{4 \pi \mathcal{D}_{L}^{2}} \left ( \frac{dl}{d\theta} \right )^{2} \frac{dl}{d\nu},
\end{equation}
where $c$ is the speed of light, $k_B$ is the Boltzmann factor, $\nu_{\rm obs}$ is the observed frequency of the redshifted line, and $dn/dM$ is the halo mass function, for which we adopt the Sheth-Tormen function \citep{Sheth:1999mn}. $L(M,z)$ is the luminosity of CO- or [CII]-luminous galaxies (as a function of host-halo mass and redshift), and ${\mathcal D}_L$ is the luminosity distance. The terms $dl/d\theta$ and $dl/d\nu$ reflect the conversion from units of comoving lengths, $l$, to those of the observed specific intensity: frequency, $\nu$, and angular size, $\theta$. The term $dl/d\theta$ is equivalent to comoving angular diameter distance, while $dl/d\nu = c(1+z)/[\nu_{\rm obs}H(z)]$, where $H(z)$ is the Hubble parameter at a given redshift.

The power spectrum consists of two primary contributions: the clustering component ($P_{\rm clust}$), which is sensitive to the distribution of objects and typically dominates on large scales, and the shot component ($P_{\rm shot}$, sometimes referred to as the Poisson component), which arises due to the discrete nature of individual galaxies and dominates on small scales. On large scales, where clustering bias can be described by a linear relation, the clustering component can be expressed as $P_{\rm clust}(k,z) =  \left[\langle T_{\rm line}\rangle\right]^2 b_{\rm line}^2(z) P_0(k,z)$, where $P_0(k,z)$ is the linear matter power spectrum and $b_{\rm line}(z)$ is the luminosity-weighted linear bias of the line emitting galaxy. This bias can be further written as
\begin{equation}\label{eq:Gbias}
b_{\rm line}(z) = \frac{\int dM  \ \frac{dn}{dM} \  b_h(M,z) L(M,z)  
}{\int dM \ \frac{dn}{dM} \ L(M,z)},
\end{equation}
with $b_h(M,z)$ being the linear halo bias, for which we adopt prediction of Sheth-Tormen mass function. The shot component of the power spectrum takes the form of 
\begin{equation}
P_{\rm shot}(z) = \frac{c^4 }{4k_B^2 \nu_{\rm obs}^4}  \int dM \frac{dn}{dM} {\left[\frac{L(M,z)}{4 \pi \mathcal D_L^2} 
\left ( \frac{dl}{d\theta} \right )^{2} \frac{dl}{d\nu} \right ]}^2.
\end{equation} 

Theory and current observational data suggest that both CO and [CII] exist in high-redshift galaxies ($z\gtrsim6$) \citep{Venemans:2015hyr, Venemans:2017dee, Popping:2016}, and can therefore be used as tracers of the growth of structure in the early Universe. However, the strength of this emission is subject to a large uncertainty, and the predicted power spectrum is very sensitive to the astrophysical modeling, which in turn can impact the constraints on PNG by more than a factor of 2 \citep{MoradinezhadDizgah:2018lac}. In our analysis presented here, we use the results of \citep{Behroozi:2012iw} to model the dependence of star formation rate on halo mass and redshift, and assume the luminosities of CO and [CII] can be written as a function of star formation rate, adopting the models of \citep{Li:2015gqa, Silva:2014ira}.

In modeling $P_{\rm clust}$, we additionally account for redshift-space distortions and the Alcock-Paczynski (AP) effect. The former is due to the fact that the power spectrum is measured in redshift-space, where peculiar velocities of galaxies distort their distribution. The latter arises from the fact that one assumes a reference cosmology to infer distances and length scales, which if incorrect, will distort the power spectrum measurement. Further details on our modeling, along with impact of modeling uncertainties on our forecasts, can be found in our accompanying work \citep{MoradinezhadDizgah:2018lac}.

Local shape non-Gaussianity, leads to a distinct, scale-dependent correction to the linear halo bias \citep{Dalal:2007cu,Matarrese:2008nc,Afshordi:2008ru}.  Consequently, the line bias given in Eq. (\ref{eq:Gbias}) receives a scale-dependent correction, $b_{\rm line}(z) \rightarrow b_{\rm line}(k,z) = b(z)+ \Delta b_{\rm line}^{\rm NG}(k,z)$, the dominant contribution of which is given by
\begin{equation}\label{eq:NGbias}
\Delta b_{\rm line }^{\rm NG}(k,z) =   \frac{6}{5} \frac{f_{\rm NL}^{\rm loc}  \delta_c [b_{\rm line}(z)-1]}{ {\mathcal M}(k,z)}, 
\end{equation}
where $\delta_c = 1.686$ is the critical linear overdensity of spherical collapse at $z=0$, and ${\mathcal M}(k,z)$ is the transfer function, relating the linear matter density fluctuations $\delta_0$ to curvature perturbations, $\delta_0({\bf k},z) = {\mathcal M}(k,z) \zeta({\bf k})$.  

On large scales, for $k\ll 0.02\ {h\,\rm Mpc}^{-1}$, the transfer function asymptotes to $k^2$, producing a strong $k^{-2}$ dependence in $\Delta b_{\rm line }^{\rm NG}$. Such a scale dependence is unlikely to be caused by other astrophysical sources; therefore, it provides a clean window to probe PNG of local shape. Over the next few years, LSS surveys will provide significantly improved constraints on $f_{\rm NL}^{\rm loc}$ by probing progressively larger volumes, utilizing the increasing strength of the signal at larger spatial scales (e.g., \citep{Giannantonio:2011ya,dePutter:2014lna,Gariazzo:2015qea,Camera:2014bwa,Alonso:2015uua,Tucci:2016hng}). Intensity mapping surveys can leverage this strategy by providing an inexpensive method for accessing faint, distant objects at higher redshifts, and thus over larger volumes. With sufficient redshift coverage, even a survey covering a small sky area can probe scales at which the enhancement in power from local PNG is significant. \\

\section{Survey design and instrumental noise} 
For the intensity mapping of CO, we consider the $J_{1\rightarrow0}$ rotational transition (with rest-frame frequency of $\nu_{\rm rest} = 115.271$ GHz), which we will refer to as CO(1-0). At the redshift range of interest, this transition is readily accessible to ground-based experiments. For our analysis, we consider a variant of the existing CO Mapping Array Pathfinder (COMAP) \citep{Li:2015gqa}. This variant, which we will refer to as COMAP-Low, is a hypothetical future lower-frequency complement to the existing instrument, designed to perform CO EoR intensity mapping experiments. With the exception of the frequency range, we generally adopt the existing parameters for COMAP as given in \citep{Li:2015gqa}. We consider an instrument utilizing a 10-m aperture with 1000 dual-polarization detectors (twice the currently planned number), with a spectral resolution of 30 MHz and coverage between [12-24] GHz, $z = [3.8-8.6]$ (versus [26-34] GHz for the current instrument).  For this instrument, we assume that the system temperature of each element scales with frequency, such that $ T_\textrm{sys}=\nu_{\textrm{obs}} $ (K/GHz) at frequencies above 20 GHz, and $T_\textrm{sys}=20$ K below. We consider a survey covering 2000 sq. degrees, an area similar to that of the Hydrogen Epoch of Reionization Array (HERA)  \citep{DeBoer:2017}, with the instrument running at 50\% duty-cycle for a period of 5 years, for a total integration time of $\tau_{\rm tot} \approx 2\times10{^4}$ hours.

For the [CII] transition ($\nu_{\rm rest } = 1900.539$ GHz), the limited transmission of the atmosphere at sub-mm wavelengths makes ground-based observations more challenging; we therefore consider a space-based instrument. The Primordial Inflation Explorer (PIXIE), designed to study inflation via polarized emission from the CMB \citep{Kogut:2011,Kogut:2016}, is ideal to probe [CII] emission from the redshift range of   $z = [0.06 - 11.7]$ (the frequency range of  $150 - 1800 {\rm\ GHz}$) \citep{Switzer:2017kkz}. PIXIE has frequency coverage between 30 GHz and 6 THz, with 400 15-GHz synthesized frequency channels. Although relatively coarse at the lowest frequencies, such an instrument has adequate resolution for the redshift range of interest for [CII]. PIXIE is purpose-designed to conduct a full-sky CMB survey, but is suitable for wide-field [CII] intensity mapping studies. For our analysis, we limit consideration to the cleanest 75\% of the sky (matching that of the proposed polarized CMB measurement).

For an intensity mapping analysis, the per-mode instrumental noise, $P_{\rm N}$, is related to the per-voxel imaging sensitivity, $\sigma_{\rm vox}$, by $P_{\rm N} = \sigma_{\rm vox}^{2}V_{\rm vox}$, where $V_{\rm vox}$ is the co-moving volume contained within a single voxel. The per-voxel sensitivity depends on instrumental configuration. For an instrument like COMAP, $\sigma_{\rm vox} = T_{\rm sys}/\sqrt{\delta\nu\tau_{\rm int}}$, where $T_{\rm sys}$ is the system temperature of the instrument, $\tau_{\rm int}$ is the total integration time per single pointing and $\delta\nu$ is the frequency resolution for a single channel. Combining these two expressions, we can further define
\begin{equation}
P_{\rm N} = \frac{T_{\rm sys}^2}{\tau_{\rm tot}N_{\rm det}}\Omega_{\rm surv}\left(\frac{dl}{d\theta}\right)^2 \frac{dl}{d\nu},
\label{eq:inst_noise_power_ext} 
\end{equation}
where $N_{\rm det}$ is the number of detectors (i.e., single polarization feeds), and $\Omega_{\rm surv}$ is the area of our survey.

\begin{figure*}[t]
\centering
\includegraphics[width=0.5\textwidth]{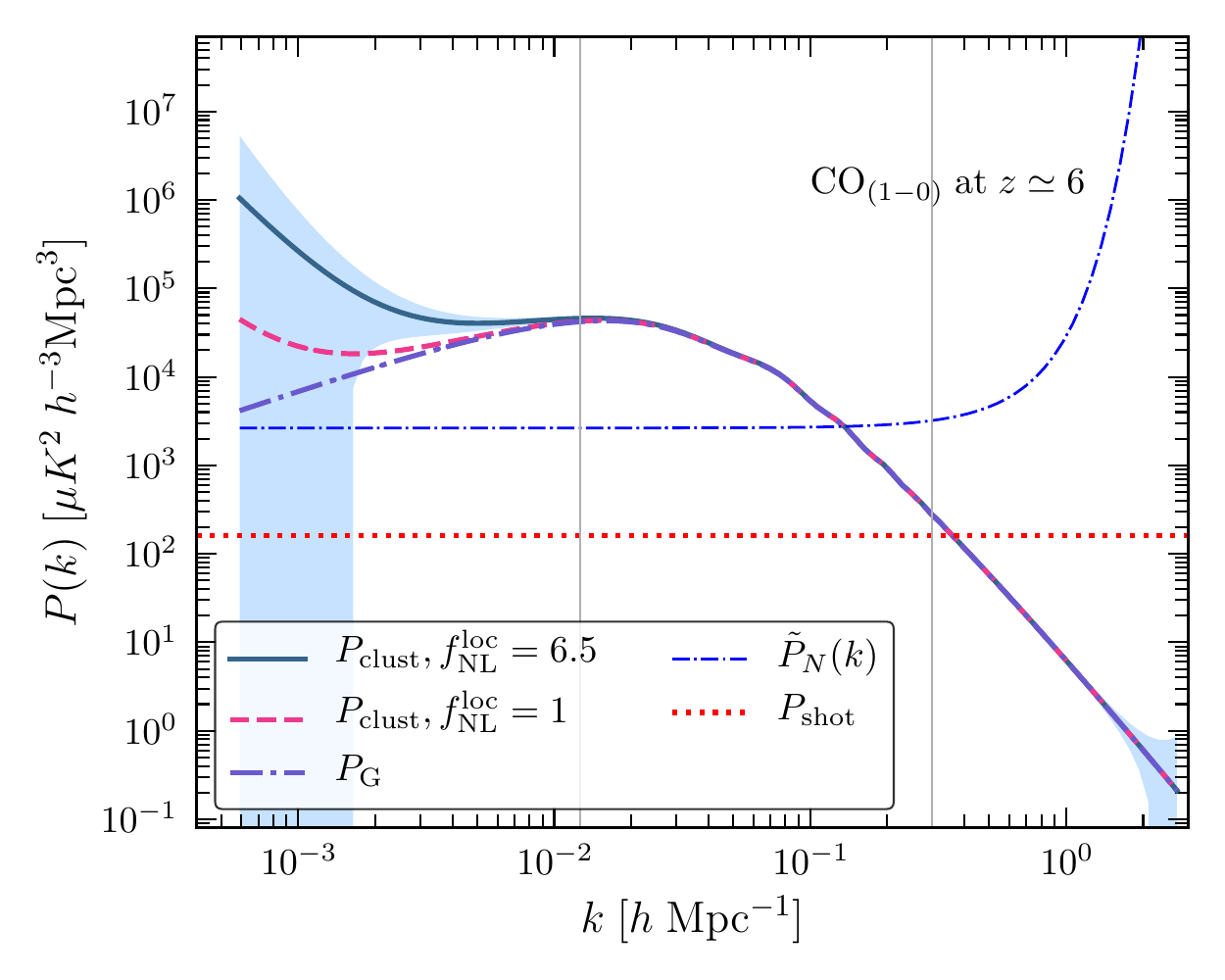}\includegraphics[width=0.5\textwidth]{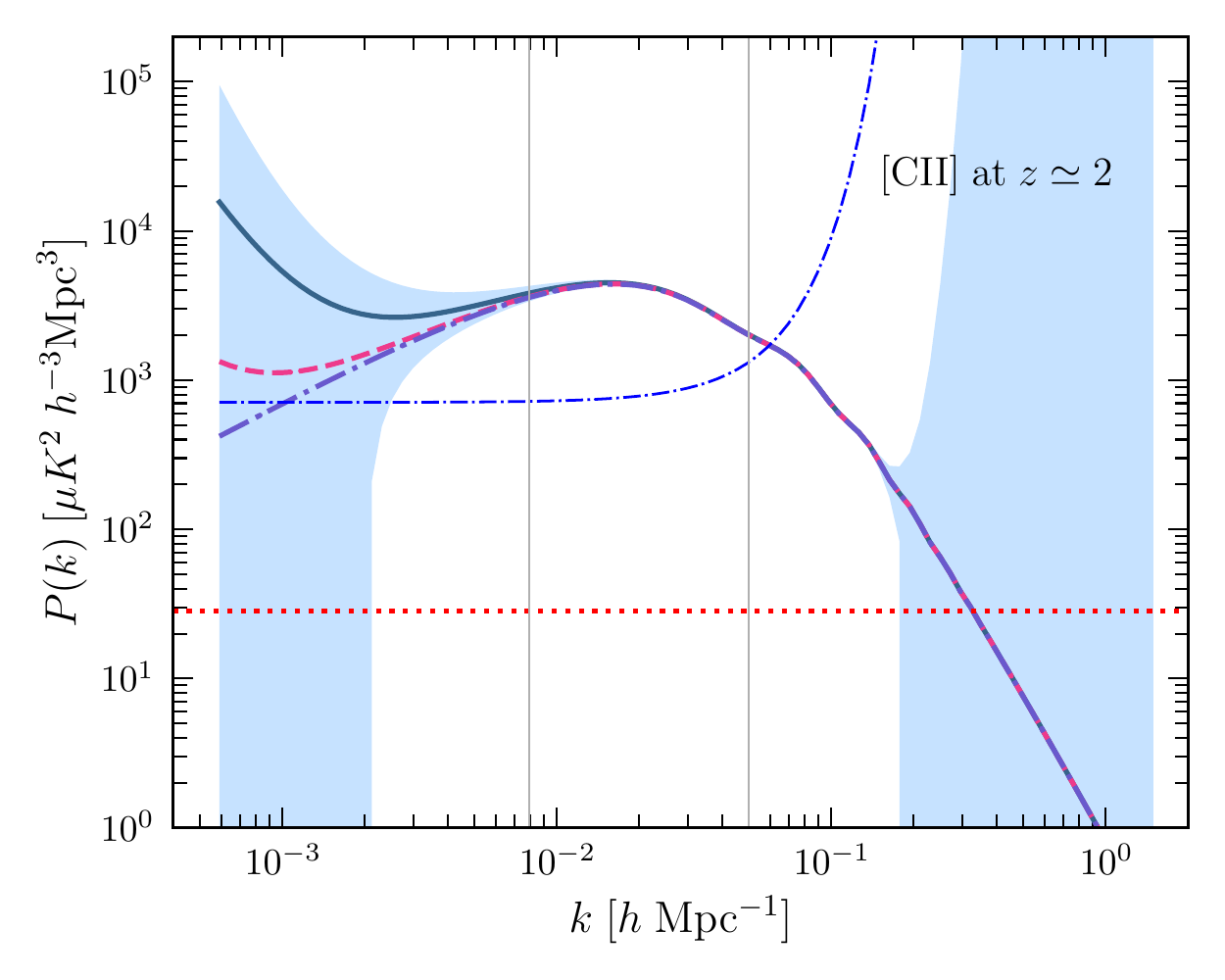}
\caption{The spherically averaged clustering component of the power spectrum   for $f_{\rm NL}^{\rm loc} = 6.5$ (solid blue) and $f_{\rm NL}^{\rm loc} = 1$ (dashed magenta) for CO(1-0) at $z\simeq 6$ (left) and [CII] (right) at $z\simeq 2$ are shown. These redshifts are chosen as they provide the tightest constraints on $f_{\rm NL}^{\rm loc}$ for the experiments considered here. The shaded blue region is the expected spherically averaged variance for the power spectrum with $f_{\rm NL}^{\rm loc} = 6.5$. For illustration, we also show the Gaussian contribution (dashed-dotted purple), instrumental noise (thin dashed-dotted blue) as well as the shot contribution (dotted red).  The vertical lines correspond to the largest, $k_{\rm min}$, and smallest scales, $k_{\rm max}$, considered in our forecast, the choice of which is described in the text.}
\label{Fig:PS}
\end{figure*}

For illustration, we show in Figure \ref{Fig:PS} the spherically averaged clustering contribution to the redshift-space line intensity power spectrum, $P_{\rm clust}(k) =  \int_{-1}^1d \mu/2 \ P_{\rm clust}(k,\mu)$, for CO and [CII], accounting for the AP effect and in the presence of PNG with $f_{\rm NL}^{\rm loc} = 6.5$.  Here  $\mu$ is the cosine of the angle with respect to the line of sight. The expected spherically averaged variance is shown as shaded blue region and is given by \citep{Lidz:2011dx},
\begin{equation}\label{eq:error}
\frac{1}{\sigma_P^2(k)} = \sum_\mu \frac{k^3V_{\rm survey}}{8\pi^2}\frac{\Delta \mu}{{\rm var} [P(k,\mu)]},
\end{equation}
where ${\rm var} [P(k,\mu)] = \left[P_{\rm clust}(k,\mu) +P_{\rm shot}(k)+ \tilde P_{\rm N} (k,\mu)\right]^2$. Here $\tilde P_{\rm N}(k,\mu) = P_{\rm N} e^{(k_{||}/k_{||,{\rm res}})^2+(k_\perp/k_{\perp,{\rm res}})^2}$, where $k_{||} = k\mu$ and $k_{\perp}^2 = k^2 - k_{||}^2$ are the components of the wavenumber parallel and perpendicular to the line of sight and $k_{||,{\rm res}}$ and $k_{\perp,{\rm res}}$ represent the finite resolution of the survey in the two directions. We adopt logarithmic bins of width $\epsilon= d \ln k$. For illustration, we also show the shot (dotted red) and the clustering components of the power spectrum with Gaussian initial conditions (dashed-dotted purple). We also show the spherically averaged instrumental noise $\tilde P_N(k) = \int_{-1}^1d \mu/2 \ \tilde P_N(k,\mu)$ (thin dashed-dotted blue).  The vertical lines correspond to the largest and smallest scales, $k_{\rm min}$ and $k_{\rm max}$, that we consider in our forecast.  For [CII], due to limited resolution of PIXIE, the value of $k_{\rm max}$ is generally set by the frequency resolution ($\delta\nu_{\rm obs}=15$ GHz). For CO, we choose $k_{\rm max}= 0.15 \ h \ {\rm Mpc}^{-1}$ at redshift zero, while at other redshifts we set it such that the variance of the density field at that redshift is the same as at $z=0$, and further impose a conservative bound of $k_{\rm max}<0.3 \ h \ {\rm Mpc}^{-1}$ to assure the validity of the assumption of linear bias. To set $k_{\rm min}$, we conservatively assume that foregrounds are only smooth over an interval of $\log_{10}[\Delta(1+z)]=0.1$ (i.e., 20\% in bandwidth at a given frequency $\nu_{\rm obs}$), such that $k_{{\rm min,}\parallel}=2\pi[(10^{0.1}-1)\nu_{\rm obs}\, dl/d\nu]^{-1}$. We also report the constraints adopting a more optimistic assumption that the foregrounds are smooth over the full frequency range of the instrument, up to $\log_{10}[\Delta(1+z)]=0.5$.

{\it Fisher forecast:} We perform a Fisher matrix analysis to forecast the potential of the CO(1-0) and [CII] intensity mapping surveys to constrain PNG. In our forecast, we vary the amplitude of the local PNG, setting its fiducial value to $f_{\rm NL}^{\rm loc}=1$. Additionally, we vary five cosmological parameters, namely, the amplitude and the spectral index of primordial fluctuations, the Hubble parameter, and the energy density of cold dark matter and baryons with the fiducial values set to best-fit parameters from Planck 2015  \citep{2016A&A...594A..13P}. We also vary the velocity dispersion (which affects the modeling of redshift-space distortions) with the fiducial value set to $\sigma_{{\rm FOG},0} = \ 250 \ {\rm km} s^{-1}$. This value is set assuming that the line is emitted from blue star-forming galaxies which are expected to reside in lower-mass halos, and hence have a low velocity-dispersion. Instead of varying the bias as a free parameter, we assume that it is given by Eq. (\ref{eq:Gbias}), which has a dependence on cosmological parameters. We bin each survey into redshift bins of approximate width $\log_{10}[\Delta(1+z)]=0.1$. 

\section{Results} 
Using the entire available three-dimensional volume, we find that the PIXIE and COMAP-Low experiments are capable of reaching $68\%$ C.L. of $\sigma(f_{\rm NL}^{\rm loc}) =3.9$ and $\sigma(f_{\rm NL}^{\rm loc}) =3.4$ respectively, imposing Planck priors on cosmological parameters. If we assume smooth foregrounds up to a maximum of two octaves in setting the value of  $k_{\rm min}$, we obtain $\sigma(f_{\rm NL}^{\rm loc}) = 1.4 $ for COMAP and  $\sigma(f_{\rm NL}^{\rm loc}) =1.7$ for PIXIE.  The strength of the constraints on local shape PNG from each survey is determined by two factors: largest scales accessible, and per-mode noise of the instrument. For PIXIE, the strongest constraints arise from $z\lesssim3$; above this redshift, the increasing per-mode noise overtakes the enhancement in power from PNG at low $k$. In contrast, the increased per-mode sensitivity of COMAP-Low provides relatively even constraints on $\sigma(f_{\rm NL}^{\rm loc})$ across all redshift bins. 

In the analysis presented here, we note that we have assumed that foregrounds, interloper lines, and systematic errors are well constrained, and not a limiting factor in our analysis. However, each one of the aforementioned effects can have a significant impact on the fidelity with which one can accurately measure the power spectrum; we therefore consider the impact of each on both of our hypothetical experiments.

\noindent\emph{Spectral Foregrounds}: The removal of interloper line emission is an area of active development. For CO, the contribution of interloper lines is likely negligible \citep{2017ApJ...846...60C}. However, lower redshift CO emission (from several rotational transitions of the molecule) could represent a significant foreground for [CII]. Existing theoretical work suggests a variety of methods for removing or reducing the impact of spectral foregrounds (e.g., \citep{Lidz:2016lub, Cheng:2016}). We note that this contamination will be most significant for $z\geq6$, where PIXIE provides limited constraints. This potential contamination is therefore unlikely to affect the results presented here, although it may limit other EoR-targeted experiments seeking to constrain $f_{\rm NL}^{\rm loc}$. 

\noindent\emph{Continuum Foregrounds}: Experiments targeting CO and [CII] are likely to benefit from decades of work in continuum foreground modeling that have been performed as part of CMB surveys, at least for the largest scale modes that are of interest for a PNG-focused measurement. To first order, such foregrounds will contaminate around $k_{z}=0$, which may cap the maximum achievable value of $k_{\rm min}$. Their impact should, however, be much smaller at higher values of $k_{z}$, and the minimum value of $k_{z}$ will depend in part on the smoothness of these foregrounds (hence the conservative and optimistic estimates) \citep{2015ApJ...814..140K,Switzer:2017kkz}. For CO, the dominant foregrounds are likely to arise from Galactic synchrotron emission and radio point sources, whereas for [CII] the dominant foregrounds are likely to arise from Galactic dust emission and the cosmic infrared background \citep{2016A&A...594A..11P}. In the absence of instrument systematics, residuals from these continuum foregrounds at the $k_{z}\ne0$ modes ought to be similar or subdominant to the shot power component, except at very low $\ell$, where Galactic continuum emission may arise as a significant contaminant \citep{Switzer:2017kkz,2015ApJ...814..140K}. Because $k_{\rm min}$ is primarily set by choice of frequency interval for the redshifts of interest, these low-$\ell$ modes do not necessarily provide access to larger-scale modes, and at worst case, can be discarded for a modest sensitivity penalty of $\sim10\%$ in the case of PIXIE, and $<5\%$ for the case of COMAP. 

\noindent\emph{Instrument Systematics}: For both CO and [CII], instrument systematics -- particularly, frequency-dependent structure in the instrument response -- can cast significant power into what would otherwise be relatively ``clean", inducing potentially scale-dependent foreground contributions to a power spectrum measurement. The degree of contamination depends largely on the degree to which frequency-based calibration errors are present in the system and how stable they are over the course of observations. Analysis of CO- and [CII]-based instruments suggests that bandpass errors of a few percent is the threshold at which such errors can become more significant \citep{2015ApJ...814..140K,Switzer:2017kkz}, making it a particularly important instrument systematic to have well constrained. One should note that static bandpass errors in the presence of an isotropic signal (i.e., instrument noise, CMB) will contaminate modes around the $k_{z}$ axis (i.e., $k_{z} \approx k_{y} \approx 0$), which may further limit access in these measurements to low $k$ modes. These issues are detailed further in our accompanying work \citep{MoradinezhadDizgah:2018lac}.

\section{Conclusions} 
Intensity mapping can provide a powerful probe of cosmology, that is highly complementary to galaxy surveys. We provide a first forecast for the potential of CO and [CII] lines to probe PNG and show that the proposed COMAP-Low and PIXIE  can achieve $68\%$ C.L. of $\sigma(f_{\rm NL}^{\rm loc}) = 3.4 $ and  $\sigma(f_{\rm NL}^{\rm loc}) =3.9$. These constraints are an improvement over those from {\it Planck}, and are comparable to those from upcoming galaxy surveys such as EUCLID \citep{2018LRR....21....2A} ( $\sigma(f_{\rm NL}^{\rm loc}) = 3.9$  \citep{2018JCAP...01..010M}), DESI \citep{2016arXiv161100036D} ($\sigma(f_{\rm NL}^{\rm loc}) = 4.8$ \citep{Gariazzo:2015qea}) and LSST \citep{2009arXiv0912.0201L} ($\sigma(f_{\rm NL}^{\rm loc}) = 1.4$ \citep{2018JCAP...01..010M}). 

The multi-tracer technique can further improve our results by minimizing the cosmic variance noise \citep{Seljak:2008xr}. For the case of PIXIE, multi-tracer analysis is particularly timely since the range of redshifts that provide the most constraining power is well-matched to several of the upcoming galaxy surveys. Neglecting  potential population covariances, we obtain $\sigma(f_{\rm NL}^{\rm loc}) = 0.96$ for PIXE+LSST. We defer more detailed analysis to future work.

We note that our analysis has focused on instruments with present prototypes or existing designs. There is significant potential for a more optimized experiment to probe the distribution of matter at large scales and high redshifts via [CII] and CO intensity mapping, offering a promising window to constrain primordial non-Gaussianity. This potential is bolstered by the complementary observational requirements for PNG-focused intensity mapping surveys and select future surveys targeting CMB and EoR-related science. At the cosmic variance limit, an EoR-focused ($z=[6-10]$) CO or [CII] intensity mapping survey would achieve $\sigma(f_{\rm NL}^{\rm loc}) \approx 0.3$, suggesting that a ground-based experiment with limited sky-coverage provides  significant constraints on $f_{\rm NL}$ \citep{MoradinezhadDizgah:2018lac}.

\section{Prospects} 
Given the potential of intensity mapping to probe PNG, understanding the dependence of the constraints on astrophysical modeling is critically important. Equally important are more detailed studies of the impact of foregrounds and systematics, particularly on the large-scale modes which are essential in constraining local PNG from power spectrum measurements. We give further consideration of these aspects, and the study of optimal experimental setup, best-fit to probe PNG with intensity mapping, in our accompanying work \citep{MoradinezhadDizgah:2018lac}.  

\section*{Acknowledgements}
It is our pleasure to thank Tzu-Ching Chang, Olivier Dor\'{e}, and Eric Switzer for fruitful discussion and helpful comments on this manuscript. We further thank Eric Switzer for providing us the updated PIXIE noise curves and Julian Mu\~noz for providing us with the {\it Planck} Fisher matrix used to obtain Planck priors.

\end{document}